\documentclass[aps,twocolumn,superscriptaddress,altaffilletter,lengthcheck,tightenlines,showpacs,showkeys]{revtex4}

\usepackage{multirow}            

\newcommand{\al}{\alpha}

\newcommand{\ben}{\begin{eqnarray}}
\newcommand{\een}{\end{eqnarray}}
\newcommand{\be}{\begin{equation}}
\newcommand{\ee}{\end{equation}}
\newcommand{\ba}{\begin{eqnarray}}
\newcommand{\ea}{\end{eqnarray}}
\newcommand{\n}{\label}

\newcommand{\la}{\lambda}
\newcommand{\ga}{\gamma}
\newcommand{\ro}{\rho}

\newcommand{\om}{\omega}

\usepackage[dvipdf]{epsfig}

\usepackage{color}

\begin{document}
\date{\today}

\title{Crossing the phantom divide with k-essence in braneworlds }

\author{Luis P. Chimento}\email{chimento@df.uba.ar}
\affiliation{Department of Theoretical Physics, University of the Basque Country,
P.O. Box 644, 48080 Bilbao, Spain \\
and IKERBASQUE, the Basque Foundation for Science, 48011, Bilbao, Spain}

\author{M\'onica  Forte}\email{forte.monica@gmail.com}
\affiliation{ Departamento de F\'{\i}sica, Facultad de Ciencias Exactas y
Naturales,  Universidad de Buenos Aires, Ciudad
Universitaria, Pabell\'on I, 1428 Buenos Aires, Argentina}

\author{Mart\'{\i}n G. Richarte}\email{martin@df.uba.ar}
\affiliation{ Departamento de F\'{\i}sica, Facultad de Ciencias Exactas y
Naturales,  Universidad de Buenos Aires, Ciudad
Universitaria, Pabell\'on I, 1428 Buenos Aires, Argentina}

\begin{abstract}

We study a flat 3-brane in presence of a linear $k$ field with nonzero cosmological constant $\Lambda_{4}$. In this model the crossing of the phantom divide (PD) occurs when the $k$-essence energy density becomes negative. We show that in the high energy regime the effective equation of state has a resemblance of a modified Chaplygin gas while in the low energy regime it becomes linear. We find a scale factor that begins from a singularity and evolves to a de Sitter stable stage while other solutions have a super-accelerated regime and end with a big rip. We use the energy conditions to show when the effective equation of state of the brane-universe crosses the PD.
\end{abstract}

\maketitle

\section{Introduction} 

In the last decades there has been a great interest in the development of the string and membrane theories. The interest of the string theory lies in the fact that it may provide a unified description of the gauge interactions and gravity. String theory predicts the existence of the so called $p$-branes, that is, $p+1$-dimensional sub-manifold of 10-dimensional spacetime on which open string ends. Gauge fields and gauge fermions which correspond to string end points can only move along these $p$-branes, while gravitons which are represented by closed strings (loops) can propagate into the full spacetime \cite{Polchinky1},\cite{Polchinky2} . These ideas offer a new perspective not only to understand the cosmological evolution of the universe but also to  address puzzling issues rooted in particle physics \cite{ADD1},\cite{ADD2},\cite{ADD3},\cite{ADD4},\cite{HW1},\cite{HW2},\cite{RSH1},\cite{RS1},\cite{RS2}. An important ingredient coming from Superstring and the M-theory is the notion of braneworlds. In this framework our observed Universe may be conceived as a 3-dimensional domain wall or 3-brane  embedded in a higher dimensional spacetime usually called the bulk \cite{RSH1}. 
Two of the most promising brane-world scenarios are the Randall- Sundrull type I-II (RS) model \cite {RS1}, \cite{RS2} and the Dvali-Gabadadze-Porrati (DGP) gravity \cite{DGP1}. On the one hand the RS type I model has a non-flat extradimension in which the universe is regarded a 3-brane located at the fixed point of ${\mbox{S}^{1}/\mbox{Z}_{2}}$ orbifold in $\mbox{AdS}_{5}$ spacetime.  Whereas  the RS type II
arises as a new alternative compactification with gravity \cite{RS2}. These models have drawn much attention because they offer a new possibility for adressing both the gauge hierarchy problem and the cosmological constant problem. On the other hand, the DGP model suggested a brane-induced model  with a flat large extra dimension in 5-dimensional Minkowski spacetime. The model  was regarded as an interesting  attempt for modifying gravity in the infrared region \cite{DGP1}, so it provides a modification of gravitational laws at large distances and allows, by means of a non trivial mechanism, to recover the 4-dimensional Einstein gravity at short distance \cite{DGP1}.

The observational data  seem to confirm the existence of an epoch of accelerated expansion in the early universe \cite{WAMP1},\cite{WAMP2},\cite{WAMP3},\cite{WAMP4},\cite{WAMP5}. In addition,  the observations of distant supernovas type Ia \cite{SNIa1},\cite{SNIa2} show that the universe at present is expanding with acceleration also. The idea that some kinds of field could be the agent driving those two periods of expansion  is widely accepted. In particular, a great amount of works have been invested in studying cosmological scenarios with non-canonical kinetic term known as $k$-essence models. The $k$-essence has its origin in Born-Infeld action of string theory \cite{BIS1},\cite{BIS2},\cite{BIS3},\cite{BIS4},\cite{BIS5},\cite{BIS6},\cite{BIS7} and it was  introduced as a possible model for inflation \cite{INFLKE1},\cite{INFLKE2}. In this context,  inflation is polelike, that is, the scale factor evolves like a negative power of the cosmic time \cite{INFLKE1},\cite{INFLKE2},\cite{SCH1},\cite{CHI1}.  Lately,  efforts in the framework of $k$-essence have been directed toward model building using power law solutions which preserve \cite{CHI1},\cite{PAD1},\cite{AFE1},\cite{CHFE1} or violate the weak energy condition \cite{ACHR}. In addition, a $k$-essence model with a divergent sound speed (called atypical $k$-essence) was carefully analyzed in  Refs. \cite{MCL},\cite{CHIR1}. In the latter it was shown that the model fix the form of the Lagrangian for $k$-essence matter. Many  others aspects concerning $k$-essence theory have been explored in the literature  such as an unifying model of  dark components with $k$-essence \cite{SCH1},\cite{BM1}, $k$-essence as dark energy source \cite{VIK}, purely kinetic multi $k$-essence model crossing the 
PD \cite{CHIR2},\cite{SSDAS}, tracking solution \cite{DKS} or holography \cite{HOLOK1},\cite{HOLOK2},\cite{HOLOK3} and so on.

The late cosmological evidences \cite{observacross1},\cite{observacross2},\cite{observacross3},\cite{observacross4}  seem to indicate that the dark energy component  might  have  crossed the PD, that is, it could have  evolved from a non-phantom phase  characterized by the equation of state $\omega>-1$ to a phantom stage with $\omega<-1$. Hence, in the light of the recent observational data which slightly favors  the crossing of the PD, $\omega=-1$, many authors started  an exhausted seek  to implement  the crossing for different kind of sources within  Einstein's gravity \cite{hotopic1},\cite{hotopic2},\cite{hotopic3},\cite{hotopic4},\cite{hotopic5},\cite{hotopic6},\cite{hotopic7}. Furthermore, the above aforementioned  topic has been analyzed in the context of modified gravity theories and brane-worlds model also (see  \cite{CruceBrane1}-\cite{CruceBrane16}). In particular, the crossing of the PD was examined in the context of  many interesting theoretical models  such as the warped DGP brane, Gauss-Bonnet gravity, LDGP model, dilatonic braneworlds, brane models with bulk matter, DGP-inspired $F(R,\phi)$-gravity (see \cite{CruceBrane1}-\cite{CruceBrane16}).

Recently, power law solutions on a 3-dimensional brane coupled to a tachyonic field were obtained\cite{SA1} by using  a well known algorithm developed  in  \cite{PAD1}.  Later on, an exact solution for a linear $k$ field was found in \cite{CHIRICHA}, showing the relevance of the $k$-essence matter on the brane. There, it showed  that  the brane equations admit an extended tachyons with negative energy density \cite{CHIRICHA}. 

In this paper we present the cosmological equations for a flat FRW brane in the presence of $k$-essence with a four dimensional cosmological constant. We examine the crossing of the PD and the energy conditions by introducing an effective description for the lineal $k$ field and the atypical $k$-essence. In the appendix we classify the solutions. Finally, the conclusions are stated.

\section{Crossing of the  PD}

We will focus on the study of 3-dimensional brane associated to a flat FRW universe with cosmological constant $\Lambda_{4}$. The brane  is filled with a $k$-essence field $\phi$. The induced metric takes the form $g_{\mu\nu}={\mbox{diag}} [-1,a^{2}\delta^{j}_{i}]$, where $a$ is the scale factor and  $H=\dot{a}/a$ is the Hubble expansion rate while the modified Einstein equations on the brane read (see \cite{BRA1}, \cite{BRA2},\cite{BRA3},\cite{BRA4},\cite{BRA5})
\be
 3H^{2}=\rho_{\phi}\Big(1+ \frac{3}{\lambda^{2}}\rho_{\phi}\Big) + \Lambda_{4} \label{fm},
\ee
\be
{\dot\rho}_{\phi}+3\,H\,(\om_{\phi}+1)\rho=0,\label{cphi1}
\ee
with $\rho_{\phi}=V[F-2xF_{x}]$ and $p_\phi=-VF$. Here, $F$ is a differentiable function of the kinetic energy $x=-\dot{\phi}^{2}$, $F_{x}=d\,F/d\,x$, $V$ is the potential and  $\om_{\phi}=p_{\phi}/\rho_{\phi}$ is the state parameter of the $k$-essence. 

The quadratic term in the energy density (\ref{fm}) gives the correction to the standard cosmology. It  dominates  at the early stage of the universe when the Hubble expansion behaves linearly with the energy density $H\propto \rho_{\phi}/\la$, modifying the dynamic of the universe for $\rho_{\phi}\gtrsim \la^2$. However, the quadratic term becomes negligible in the limit $\la\rightarrow \infty$ recovering the four-dimensional general relativity $H\propto \sqrt{\rho_{\phi}+\Lambda_{4}}$.

Many authors have explored the crossing of the PD in brane-world cosmologies by using different sources or modified gravity theories \cite{CruceBrane1}-\cite{CruceBrane16}. Here, we will investigate the crossing of the PD, which appears to be compatible with the  recent cosmological observations, in a RS type II model with $k$-essence. Let us consider a $k$-essence fulfilling the requisites $\rho_\phi=\alpha_0 V$ with $\alpha_0=cte$ and a $k$ field evolving linearly with the cosmological time, $\phi=\phi_{0}t$ with $\phi_{0}=cte$. In the case of  $\om_{\phi}=\om_{0}$ it is possible to show that the conservation equation (\ref{cphi1}) can be easily integrated  to obtain  the energy density $\ro_\phi=\ro_0 a^{-3(\om_{0}+1)}$ as well  the potential $V=\rho_{0}a^{-3(\om_{0}+1)}/\alpha_0$. Besides, the equation of state for this $k$-essence is $p_\phi=\om_{0}\ro_\phi$. Now, we associate the brane equations to an effective Friedmann cosmology introducing an effective fluid as follows
\be
\label{deff}
3H^{2}=\rho_{eff}, \quad -2\dot{H}=\rho_{eff} + p_{eff}.
\ee
Combining these equations with the effective state parameter $\om_{eff}=-1-2\dot{H}/3H^{2}$, we obtain 
\begin{equation}
\om_{eff}=-1+ \frac{(\om_{0}+1)\left[1+ \frac{6}{\lambda^{2}} \ro_{0} a^{-3(\om_{0}+1)}\right]} {1+ \frac{3}{\lambda^{2}}\ro_{0} a^{-3(\om_{0}+1)} + \frac{\Lambda_{4}}{\ro_{0}} a^{3(\om_{0}+1)} }.\label{Geffa}
\end{equation}
Near the initial singularity the effective fluid has a linear equation of state $p_{eff}\approx (2\om_{0}+1)\ro_{eff}$ representing dust matter for $\om_{0}=1/2$. However, at late stage the effective fluid describes a final de Sitter-like phase. For $\ro_{0}<0$, we obtain the crossing of the PD, $\omega_{eff}=-1$, in the RS type II model (see Fig.1). The same occurs for $\om_{0}<0$ as it can be seen from Eq. (\ref{Geffa}). Recently, in the context of the standard cosmology \cite{CACHI}, it showed the crossing of the PD with a single massive classical Dirac field. In our brane model this case it corresponds to the values $\ro_{0}<0$ and $\om_{0}=1$. Hence, a classical Dirac field localized on the brane allows to cross the PD while the non-zero $\Lambda_{4}$ drives an accelerated expansion at late time.

\begin{figure}[!ht]

\includegraphics[height=6cm, width=8.2cm]{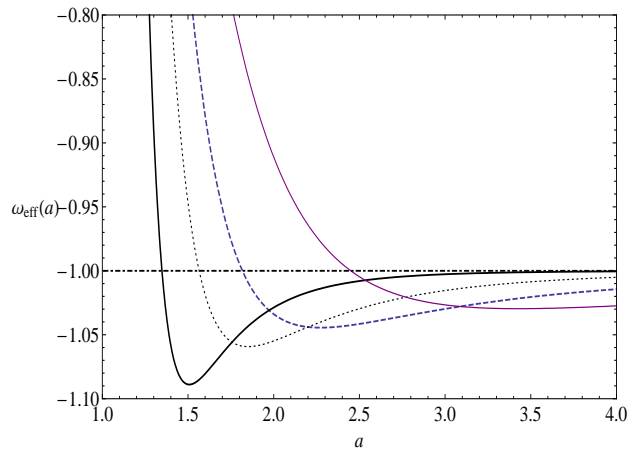}

\caption{We plot $\omega_{eff}$ with $\ro_{0}<0$ for  $\om_{0}=-1/3,0,1/3,1$. The PD corresponds to the line $\omega_{eff}=-1$.}.

\end{figure}

To complement the above study, we have found the scale factor to analyze the evolution of the braneworlds and show different cases where the crossing of the PD is achieved (see the classification in the appendix). Here we  will focus on the initial $a_{i}$ and final $a_{f}$ behavior of the scale factor, and give a qualitative description of the evolution of the 3-brane when $12\Lambda_{4}>\la^2$.

\vskip .2cm
$\triangleright$ Type A:~$\rho_{0}<0, \om_{0}>-1$.
\vskip .2cm 

The universe begins at a singularity as $a_i\approx {t}^{1/3(\om_{0}+1)}$, and ends in a de Sitter stage  as $a_{f}\approx e^{\sqrt{\Lambda_{4}/3}t}$.  At early times the effective state parameter varies over the interval $\om_{i}\in(-1, 3)$, later on, it crosses the PD and  asymptotically it  goes to $\om_{f}=-1$ (see Fig.2a). The potential interpolates between $V\approx |\la|(\phi/\phi_{0})^{-1}$, at early times and $V\approx V_{0} e^{-3(\om_{0}+1)\sqrt{\Lambda_{4}/3}(\phi/\phi_{0})}$ at late times. Then the solution turns out to be stable because the potential approaches its minimum value for large times.

\vskip .2cm
$\triangleright$ Type B:~$\rho_{0}<0, \om_{0}<-1$.
\vskip .2cm

The  universe begins in the distant past at $t=-\infty$, as $a_i\propto e^{\sqrt{\Lambda_{4}/3} \Delta t}$ and evolves toward to a big rip at $t=0$, as $a_f\approx \Delta t^{-1/3|(\om_{0}+1)|}$ passing through an intermediate stage of super-acceleration where it crosses the PD (see Fig.2b). On the other hand, the $k$ field is driven by a potential which interpolates between $V\propto  e^{-3\sqrt{\Lambda_{4}/3} |(\om_{0}+1)\phi/\phi_{0}|}$ and $V\approx |\la|(\phi/\phi_{0})^{-1}$.

\begin{figure}[!ht]
\includegraphics[height=9cm, width=7.5cm]{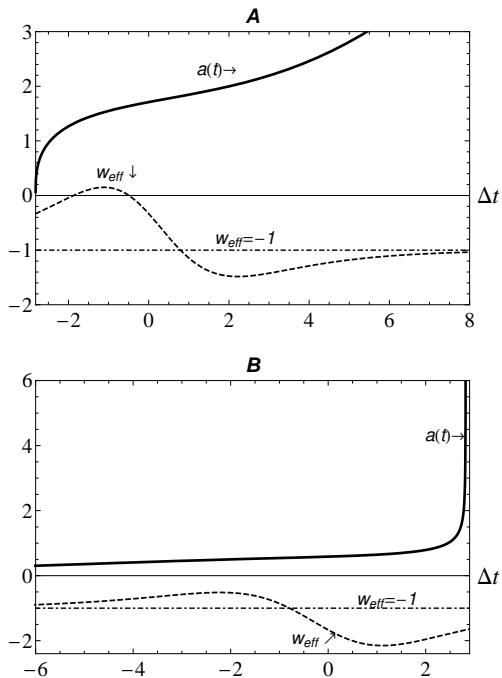}
\caption{ We plot the scale factor and the $\omega_{eff}$ as functions of the cosmic time for universes which cross the PD.}.
\end{figure}
From the theoretical point of view the expanding solutions, types A and B, seem reasonable and interesting enough to keep them in mind \cite{BRS1}-\cite{BRS6}. Later we will examine the energy conditions for these type of solutions. 

In the low  energy regime we obtain the expected linear equation of state, namely, $p_{eff} \approx - (\om_{0}+1)\Lambda_{4}+\om_{0}\rho_{eff}$, whereas in the high energy regime we get the nonlinear equation of state
\be
\n{nles}
p_{eff}\approx (2\om_{0}+1)\ro_{eff}\pm\frac{(\om_{0}+1)\la}{\sqrt{3}}\,\ro_{eff}^{1/2}.
\ee  
It has the form of the enlarged Chaplygin gas $p_{eff}=B\ro_{eff}+ C\ro^{-\nu}_{eff}$ with  $\nu=-1/2$. The deviation of the linear equation of state comes from the framework of the brane-world cosmology. A similar result was obtained in  \cite{CHIRICHA} for the vanishing  cosmological constant case. 

\subsection{Atypical $k$-essence}

We now turn our attention to the attractive atypical $k$-essence model  associated to the kinetic function \cite{CHI1},\cite{CHFE1}, \cite{CHIR1},
\begin{equation} 
\label{fdiv}
F^\infty={\al} + \beta\sqrt{-x},
\end{equation}
where ${\al}$ and $\beta$ are arbitrary constants. This model is linked with a divergent sound velocity \cite{MCL},\cite{CHIR1} and with the extended tachyon model considered in \cite{CHI1}. 
There one finds $H=\alpha V'/3\beta V$ and the $k$-essence energy density is $\rho_\phi^\infty={\al}V$, so replacing the latter expressions into the Eq. (\ref{fm}) we get the potential
\be
V(\phi)={V_{0}\over A\cosh \theta+ B\sinh\theta-\frac{\alpha}{2\Lambda_{4}}\label{Vaty}},
\ee
where the two integration constants $A$ and $B$ fulfill the constraint
\be
B^{2}-A^{2}= \frac{\alpha^{2}}{4\Lambda^{2}_{4}}\Big(\frac{12\Lambda_{4}}{\la^{2}}-1\Big),\label{vinc2}
\ee
and $\theta=\phi\sqrt{3}\beta/\al$. Choosing  $A=\alpha/2\Lambda_{4}$ and $B=\sqrt{3/\Lambda_4}\alpha/\la$, and taking $\Lambda_{4}\rightarrow 0$ we recover the inverse quadratic polynomial potential \cite{CHIRICHA}. This potential seems to extend the accustomed pole-like behavior, $V\propto \phi^{-2}$ usually found in brane cosmology with a quintessence field \cite{QUINTABRANE1}-\cite{QUINTABRANE5}. However, we need some additional information, when  we consider the Friedmann cosmology coupled to the atypical $k$-essence, to know the time dependence of $a$ or $\phi$. Hence, for a given potential, it means that we need to know one of them to get the other one.

\section{Energy Conditions and the Crossing of the PD}

Here we shall extend the investigations, related to the crossing of the PD, for the k-essence matter localized on the brane. In other words, we are interested in the modifications introduced by the quadratic energy density term in the energy conditions.

The weak energy condition (WEC) on the brane states that $\ro_{eff}\geq0$ and $\ro_{eff}+p_{eff}\geq0$. Clearly, the first requisite is satisfied when the $k$-essence energy density is positive. However, we still have another option which corresponds  to a cosmological scenario where the k-essence energy density is negative but the effective energy density turns out to be positive definite; this novel fact is  related with the quadratic term in $\rho_\phi$. At this point, the WEC can be summarized in the following two inequalities, namely,
\be
\mbox{WEC-a}~~~~~\rho_\phi\left(1+\frac{3\rho_\phi}{\la^{2}}\right) +\Lambda_{4}\geq 0,\label{wec1}
\ee
\be
\mbox{WEC-b} ~~~~~\left(\rho_\phi+p_\phi\right)\left(1+\frac{6\rho_\phi}{\la^{2}}\right)\geq 0.\label{wec2}
\ee
Then one finds that the effective energy density is positive or zero in the range $(-\infty,\rho_\phi^{-}]\cup[\rho_\phi^{+},+\infty)$,  where $\rho_\phi^{\pm}$ are given by
\begin{equation}
\n{ro}
\rho_\phi^{\pm}=\frac{\la^2}{6}\left[-1\pm\sqrt{1-12\frac{\Lambda_{4}}{\la^2}}\right],
\qquad \la^{2}\geq 12\Lambda_{4}.
\end{equation}
It corresponds to the solutions types II and III (see appendix). In the case with $\la^{2}< 12\Lambda_{4}$ the effective energy density is positive for any $k$-essence energy density and this option corresponds to the solution type I (see Appendix). 

The second condition of the WEC can be factorized  into the  product of two distinct terms,

\be
\rho_{\phi}+p_{\phi}>0 \qquad  \text{and}~~~ \left(\la^{2}+6\rho_{\phi}\right)>0,\label{ca1} 
\ee
\be
\rho_{\phi}+p_{\phi}<0 \qquad  \text{and}~~~ \left(\la^{2}+6\rho_{\phi}\right)<0.\label{ca2}
\ee
Although the sum of the pressure and energy density of the $k$-essence can be negative the WEC still is fulfilled if the $k$-essence energy density satisfies the relation:$(\la^{2}+6\rho_{\phi})<0$. As we said, this result is only possible within the brane cosmology due to the quadratic term.  It was pointed out in  \cite{CHIRICHA} the quadratic correction  allows to introduce the extended tachyons with negative energy.  

The null energy condition (NEC) in braneworlds establishes that $\ro_{eff}+p_{eff}\geq0$. It is implemented in two different manners as we have already  discussed (see Eqs. (\ref{ca1}-\ref{ca2})). These inequalities are plotted in the plane $(\rho,p)$ as functions of $\la$ and $\Lambda_{4}$ in Fig. 3a. There we show the regions where these conditions are satisfied for the solutions type I.

The the strong energy condition (SEC) on the brane states that  $\rho_{eff}+p_{eff}>0$ and $\ro_{eff}+3p_{eff}>0$.  The universe accelerates when the SEC is violated for at least one of the matter components, so that any of  the following inequalities are no longer valid   
\be
\mbox{SEC-a}\qquad \left(\rho_\phi+p_\phi\right)\left(1+\frac{6\rho_\phi}{\la^{2}}\right)\geq 0,\label{sec1}
\ee
\be
\mbox{SEC-b}\qquad \left(\rho_\phi+3p_\phi\right)+\frac{6\rho_\phi}{\la^{2}}\left(2\rho_\phi+3p_\phi\right)\geq 2\Lambda_{4},\label{ca3}
\ee

In general relativity the SEC is violated if the first term  in Eq. (\ref{ca3}) becomes negative, however in the brane-world model the things are very different. In fact, at high energy the SEC is violated when the second term in (\ref{ca3}) becomes negative.  To verify the low and high energy limits  we consider the $k$-essence fluid previously introduced with equation of state $p_{\phi}=\om_{0} \rho_{\phi}$. So, in the limit $\lambda\rightarrow\infty$,  with $\Lambda_{4}=0$, we recover the condition  $\om_{0}<-1/3$, but at high energy regime we obtain $\om_{0}<-2/3$. In Fig.3b we plot the two conditions (\ref{sec1}-\ref{ca3}) in order to see where the SEC is violated. Other aspects of the 3-brane in Randall-Sundrum type II model dominated by a phantom fluid and their connections with accelerated/decelerated stages were examined in \cite{ADBrane1},\cite{ADBrane2}.

\begin{figure}[!ht]
\includegraphics[height=5cm, width=9cm]{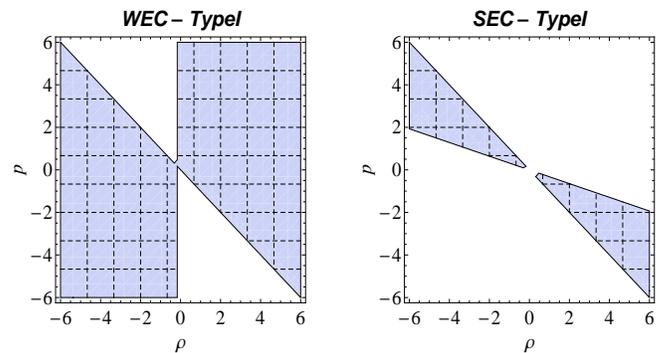}
\caption{ We plot the fulfilling of WEC and the violation of SEC for brane-world type I for  given values of $\la$ and $\Lambda_{4}$.}.

\end{figure}

To relate the energy conditions with the crossing of the PD line we recast  the energy condition as follows
\be
\mbox{WEC-b}\qquad \rho_{eff}+ p_{eff}=3H^{2}\big(\om_{eff}+1\big)\geq 0,~\label{wecb}
\ee
\be
\mbox{SEC-b}\qquad \rho_{eff}+ 3p_{eff}=3H^{2}\big(3\om_{eff}+1\big)\geq 0.~\label{secb}
\ee

For the universes types A and B the crossing of the PD, $\om_{eff}=-1$, is reached when the WEC Eq. (\ref{wecb}) is violated. In Fig.4 we have plotted the energy conditions as well as the state parameter
$\om_{eff}$ in term of the cosmic time. On the other hand, when the condition $3\om_{eff}\leq -1$ holds it leads to an accelerated and super-accelerated expansion of the 3-brane, in the same way that the universes type
A and type B do respectively(see Fig. 4). 

\begin{figure}[!ht]
\includegraphics[height=9.5cm, width=7cm]{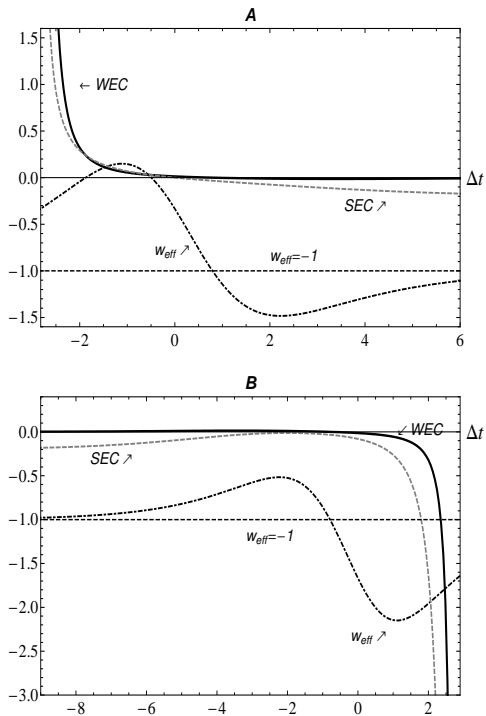}
\caption{We plot $\om_{eff}$, WEC and SEC in term of the cosmic time for  given values of $\la$ and $\Lambda_{4}$. The universes A and B belong to the solution type I.}.
\end{figure}

\section{Conclusions}

We have studied a spatially flat FRW 3-brane filled with a $k$ field evolving linearly with the cosmic time and with 4-dimensional cosmological constant. The crossing of the PD $\om_{eff}=-1$ is possible if the $k$-essence energy density is negative definite. For $\rho_{0}<0$ and $\om_{0}=1$ the $k$-essence mimics a massive classical Dirac field localized on the brane, reproducing the results obtained in \cite{CACHI}. In the high energy regime we  found that the effective equation of state has a resemblance of the  modified Chaplygin gas whereas in the low energy limit describes a linear equation of state. We  analyzed the atypical $k$-essence model and obtained agreement with the results found for the extended tachyon field \cite{CHIRICHA}. Also, we have investigated the energy conditions on the brane and established that the crossing of the PD is possible if the WEC is violated. In the appendix we have given a complete classification of the three types of families of solutions for the scale factor and the potential. 

In future researches we will analyze the role of the $k$-essence within different braneworlds set ups 
such as RS with negative tension and  bulk brane viscosity, Cardasian model, Loop Quantum Cosmology  and DGP cosmology  \cite{LPSeta1}-\cite{LPSeta5}. 

\acknowledgments

The authors thank the University of Buenos Aires for the partial support of this work during their different stages under Project No. X044, and the Consejo Nacional de Investigaciones Cient\'{\i}ficas y T\' ecnicas (CONICET)under Project PIP 114-200801-00328. MGR is supported by CONICET.


\section{Appendix}

We will find and classify the solutions of the modified Friedmann equation (\ref{fm})
\be
\label{fmadet}
3{\dot{y}}^{2}=\Lambda_{4} y^{2}+ y\ro_{0}  +\frac{3\ro^{2}_{0}}{\la^{2}},
\ee
where $y=a^{3(\omega_{0}+1)}$, $\ro_{0}=V_{0}{\alpha}_{0}$ and the dot means differentiation with respect to $\tau=3(\omega_{0}+1) t$. The scale factor reads
\be
a=\left(A\cosh b\Delta\tau + B\sinh  b\Delta\tau -\frac{\rho_{0}}{2\Lambda_{4}}\right)^{1/3(\omega_{0}+1)},\label{adet}
\ee
\be
B^{2}-A^{2}=\frac{\rho^{2}_{0}}{4\Lambda^{2}_{4}}\left(\frac{12\Lambda_{4}}{\la^{2}}-1\right),\label{vinc}
\ee
where $A, B$ are constants fulfilling Eq. (\ref{vinc}), $\Delta\tau=\tau-\tau_0$ and  $b=\sqrt{\Lambda_{4}/3}$.  For $\Lambda_{4}>0$ the solution splits into three different families: for $12\Lambda_4>\lambda^2 $ (type I), for $12\Lambda_4<\lambda^2$ (type II) and for $12\Lambda_4=\lambda^2$ (type III), they are: 
\be
a_{I}=\left[ \epsilon\frac{|\ro_{0}|}{2\Lambda_{4}}\left[\frac{12\Lambda_{4}}{\la^2}-1\right]^{1/2}\sinh 3(\omega_{0}+1)b\Delta t -\frac{\ro_{0}}{2\Lambda_{4}}\right]^{1/3(\omega_{0}+1)},\,\,\,\,\,\,\,\,\,\label{1Sol}
\ee
\be
a_{II}=\left[ \epsilon\frac{|\ro_{0}|}{2\Lambda_{4}}\left[1-\frac{12\Lambda_{4}}{\la^2}\right]^{1/2}\cosh 3(\omega_{0}+1)b\Delta t -\frac{\ro_{0}}{2\Lambda_{4}}\right]^{1/3(\omega_{0}+1)},\,\,\,\,\,\,\,\,\label{2Sol}
\ee
\be
a_{III}=\left( \epsilon  e^{ 3(\omega_{0}+1)~\sqrt{\frac{\la^2}{36}}~\Delta t} -\frac{6\ro_{0}}{\la^2}\right)^{1/3(\omega_{0}+1)},\,\,\,\,\,\,\,\,\,\,\,\,\,\,\,\,\,\,\,\,\,\,\,\,\,\,\,\,\,\,\,\,\,\,\,\label{3Sol}
\ee
with $\Delta t= t-t_{0}$ and $\epsilon=\pm 1$. Inserting  $a(t)=a(t(\phi))$ into the potential $V=V_{0}a^{-3(\omega_{0}+1)}$, we get
\be
V_{I}={V_{0}\over {\epsilon\frac{|\ro_{0}|}{2\Lambda_{4}}\left(\frac{12\Lambda_{4}}{\la^2}-1\right)^{1/2}\sinh 3(\omega_{0}+1)b\frac{\phi}{\phi_{0}} -\frac{\rho_{0}}{2\Lambda_{4}}}},\label{V1Sol}
\ee
\be
V_{II}={V_{0}\over\epsilon\frac{|\ro_{0}|}{2\Lambda_{4}}\left(1-\frac{12\Lambda_{4}}{\la^2}\right)^{1/2}\cosh 3(\omega_{0}+1)b\frac{\phi}{\phi_{0}} -\frac{\rho_{0}}{2\Lambda_{4}}},\label{V2Sol}
\ee
\be
V_{III}={V_{0}\over\epsilon  e^{ 3(\omega_{0}+1)~\sqrt{\frac{\la^2}{36}}~\frac{\phi}{\phi_{0}}} -\frac{6\ro_{0}}{ \la^2}}.\,\,\,\,\,\,\,\,\,\,\,\,\,\,\,\,\,\,\,\,\,\,\,\,\,\,\,\,\,\,\,\,\,\,\,\,\,\,\,\,\,\,\,\,\label{V3Sol}
\ee
Now, we examine the main features of these families of solutions.

For $\rho_{0}<0$ and $\omega_{0}>-1$, the scale factor type I exhibits a de Sitter regime in the distant past contracting toward a big crunch. For $\rho_{0}<0$ and $\omega_{0}<-1$, the universe contracts from the initial singularity, ending in the remote future with a vanishing scale factor. In these examples the PD, $\om_{eff}=-1$, is crossed and the WEC is violated (see Fig. 5). 

\begin{figure}[!ht]
\includegraphics[height=8cm, width=6.5cm]{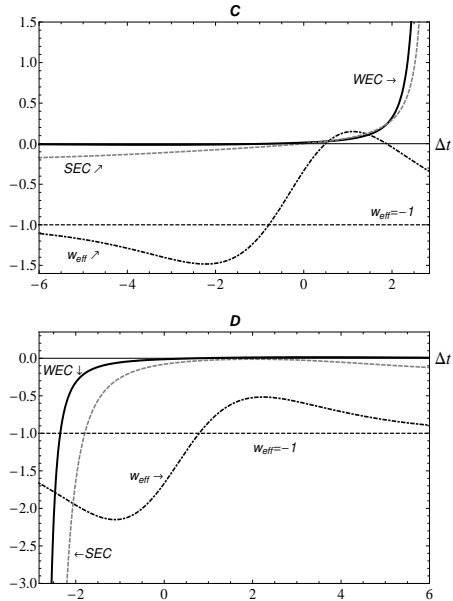}
\caption{We plot $\om_{eff}$, WEC and SEC for the time reversal solution of the universes A and B in term of the cosmic time $\Delta t$. }.
\end{figure}

For $\rho_{0}>0$ and $\omega_{0}<-1$ the scale factor type II represents a universe that ends in a big rip. 
For $\rho_{0}<0$  and $\omega_{0}<-1$ we have a bouncing universe with a finite time span and a minimum  $a_{min}=(\la^{2}/3|\rho_{0}|)^{1/3|(\omega_{0}+1)|}$ where it bounces and ends with a final big rip.
Also one finds a nonsingular universe with vanishing scale factors for the limits $\Delta t \rightarrow \pm \infty$. It represents  a universe free of singularities (see table II). 

The family type III of solutions describes singular and nonsingular universes with a de Sitter phase as well as contracting universe which begins with a constant value in the past and ends with $a=0$.

Finally, the analysis of the solutions is completed by making the time reversal of the scale factor types I, II and III. Tables I, II, and III resume all the possibles universes for the cases I, II and III respectively.

\begin{center}
\begin{table*}[!ht]
\caption{
It shows the universes I and it indicates in the first column if they cross or no the PD. $\ga_0=\omega_0+1$.}\label{etiqueta}
\vspace{5mm}
\begin{tabular}{|l|l|l|l|}\hline
Description-Crossing of the PD& $a_{i}-$$a_{f}$& $w^{eff}_{i}-$$w^{eff}_{f}$ & ($\epsilon,\rho_{0},\gamma_{0}$) \\
\hline\hline
singular-dS/No & $a_{i}=0, a_{f}\rightarrow\infty$  & $\om_{i}\in(-1,0)$, $\om_{f}\rightarrow -1$ & $(+,+,+)$\\ \hline 
\hline\hline
dS-contraction/No &$a_{i}\rightarrow\infty,a_{f}=0$ & $\om_{i}\rightarrow -1$,$\om_{f}\in(-1,0)$ & $~(-,+,+)$\\\hline 
\hline\hline
big rip/No & $a_{i}\rightarrow 0,a_{f}\rightarrow\infty$ & $\om_{i}\rightarrow -1$,$\om_{f}\in(-1,0)$ & $(+,+,-)$\\\hline 
\hline\hline 
contraction/No & $a_{i}\rightarrow\infty,a_{f}\rightarrow 0$ & $\om_{i}<-1$, $\om_{f}\rightarrow -1$ & $(-,+,-)$\\\hline 
\hline\hline
singular-dS/Yes & $a_{i}=0,a_{f}\rightarrow \infty$ & $\om_{i}\in(-1,0)$, $\om_{f}\rightarrow -1$ & $(+,-,+)$\\\hline 
\hline\hline
dS-contraction/Yes & $a_{i}\rightarrow \infty,a_{f}=0$ & $\om_{i}<-1$, $\om_{f}\in(-1,0)$ & $(-,-,+)$\\\hline 
\hline\hline
big rip/Yes & $a_{i}\rightarrow 0,a_{f}\rightarrow \infty$ & $\om_{i}\rightarrow -1$,$\om_{f}\in(-1,0)$ & $(+,-,-)$\\\hline 
\hline\hline
contraction/Yes &$a_{i}\rightarrow \infty,a_{f}\rightarrow0$ & $\om_{i}<-1$, $\om_{f}\rightarrow -1$ & $(-,-,-)$\\\hline 
\hline\hline
\end{tabular}
\end{table*}
\end{center}

\begin{center}
\begin{table*}[!ht]
\caption{
It shows the universes type II and it indicates in the first column if they cross or no the PD. $\ga_0=\omega_0+1$.}\label{etiqueta2}
\vspace{5mm}
\begin{tabular}{|l|l|l|l|}\hline
Description-Crossing of the PD& $a_{i}-$$a_{f}$& $w^{eff}_{i}-$$w^{eff}_{f}$ & ($\epsilon,\rho_{0},\gamma_{0}$) \\
\hline\hline
singular-superdS/No & $a_{i}=0, a_{f}\rightarrow\infty$ & $\om_{i}\in(-1,0)$, $\om_{f}\rightarrow -1$ &$(+,-,+)$\\ \hline 
\hline\hline
big rip/No & $a_{i}\longrightarrow0,a_{f}\rightarrow\infty$ & $\om_{i}\rightarrow -1$,$\om_{f}<-1$ & $(+,+,-)$\\\hline 
\hline\hline 
bounce-minimum/No & $a_{i}\rightarrow\infty,a_{f}\rightarrow\infty$ & $\om_{i}\rightarrow-1$, $\om_{f}\rightarrow -1$ & $(+,-,+)$\\\hline 
\hline\hline
big crunch finite time/ No & $a_{i}=0,a_{f}=0$ & $\om_{i}\in(-1,0)$, $\om_{f}\in(-1,0)$ & $(-,-,+)$\\\hline 
\hline\hline
maximum-without singularity/No & $a_{i}\rightarrow0,a_{f}\rightarrow0$ & $\om_{i}\rightarrow-1$, $\om_{f}\rightarrow-1$ & $(+,-,-)$\\\hline 
\hline\hline
bounce-big rip-finite time/No & $a_{i}\rightarrow\infty,a_{f}\rightarrow \infty$ & $\om_{i}< -1$,$\om_{f}<-1$ & $(-,-,-)$\\\hline 
\hline\hline
\end{tabular}
\end{table*}
\end{center}

\begin{center}
\begin{table*}[!ht]
\caption{
It shows the universes type III and it indicates in the first column if they cross or no the PD. $\ga_0=\omega_0+1$.}\label{etiqueta3}
\vspace{5mm}
\begin{tabular}{|l|l|l|l|}\hline
Description-Crossing of the PD& $a_{i}-$$a_{f}$& $w^{eff}_{i}-$$w^{eff}_{f}$ & ($\epsilon,\rho_{0},\gamma_{0}$) \\
\hline\hline
singular-dS/No & $a_{i}=0, a_{f}\rightarrow\infty$~ & $\om_{i}\in(-1,0)$, $\om_{f}\rightarrow -1$ & $(+,+,+)$\\ \hline 
\hline\hline
big rip/No &$a_{i}\rightarrow 0,a_{f}\rightarrow\infty$ & $\om_{i}\rightarrow-1$, $\om_{f}< -1$ & $(+,+,-)$\\\hline 
\hline\hline 
dS-ends at $a_{0}>0$ /No & $a_{i}\rightarrow 0,a_{f}\rightarrow a_{0}>0$ & $\om_{i}\rightarrow-1$, $\om_{f}\rightarrow +\infty$ & $(+,-,-)$\\\hline 
\hline\hline
starts at $a_{0}>0$-dS/No & $a_{i}\rightarrow a_{0} >0,a_{f}\rightarrow\infty$ & $\om_{i}\rightarrow-\infty$, $\om_{f}\rightarrow -1$ & $(+,-,+)$\\\hline 
\hline\hline
start with $a_{0}>0$- big crunch/No & $a_{i}\rightarrow a_{0}>0,a_{f}=0$ & $\om_{i}\rightarrow+\infty$, $\om_{f}\in(-1,0)$ & $(-,-,+)$\\\hline 
\hline\hline
contraction-end with $a_{0}>0$/No & $a_{i}\rightarrow\infty,a_{f}\rightarrow a_{0}>0$ & $\om_{i}< -1$,$\om_{f}\rightarrow+\infty$  & $(-,-,-)$\\\hline 
\hline\hline
\end{tabular}
\end{table*}
\end{center}

\end{document}